\documentclass[nohyper,12pt,letterpaper]{JHEP3}

\usepackage{amsfonts,amsmath,amsopn,amssymb,amsthm,bbm,latexsym,mathrsfs,pstricks,verbatim}
\usepackage[dvipdfm,dvips]{graphicx}

\title{A Pyramid Scheme for Particle Physics}

\author{T.\,Banks\\
Department of Physics and SCIPP\\
University of California, Santa Cruz, CA 95064\\
{\it and}\\
Department of Physics and NHETC\\
Rutgers University, Piscataway, NJ 08854\\
E-mail: \email{banks@scipp.ucsc.edu}}

\author{J.-F.\,Fortin\\
Department of Physics and NHETC\\
Rutgers University, Piscataway, NJ 08854\\
E-mail: \email{jffor27@physics.rutgers.edu}}

\abstract{We introduce a new model, the Pyramid Scheme, of direct mediation of SUSY breaking, which is compatible with the idea of Cosmological SUSY Breaking (CSB).  It uses the {\it trinification} scheme of grand unification and avoids problems with Landau poles in standard model gauge couplings.  It also avoids problems, which have recently come to light \cite{tbhh}, associated with rapid stellar cooling due to emission of the pseudo Nambu-Goldstone Boson (PNGB) of spontaneously broken hidden sector ``baryon" number.  With a certain pattern of $R$-symmetry breaking masses, a pattern more or less required by CSB, the Pyramid Scheme leads to a dark matter candidate that decays predominantly into leptons, with cross sections compatible with a variety of recent observations \cite{ambisigs}.  The dark matter particle is not a thermal WIMP but a particle with new strong interactions, produced in the late decay of some other scalar, perhaps the superpartner of the QCD axion, with a reheat temperature in the TeV range.  This is compatible with a variety of scenarios for baryogenesis, including some novel ones which exploit specific features of the Pyramid Scheme.}

\received{} \accepted{}
\preprint{RUNHETC-2009-30, SCIPP-09/04}
\begin{document}

\section{Introduction}

Direct gauge mediation models are attractive from a variety of points of view.  They are the most straightforward solution to the SUSY flavor problem of the MSSM.  The general structure of a direct gauge mediation model is that of a supersymmetric quivering moose with gauge group $G\times SU(1,2,3)$.  There are chiral fields $F_i^A$ which transform in irreducible representations of both groups, possibly including singlets which can couple to the non-singlets in the cubic superpotential.  The fields that are singlets under $G$, but not $SU(1,2,3)$, are assumed to be precisely the $3$ generations plus two Higgs fields of the MSSM.  At the scale $\Lambda_G$ the $G$ gauge interactions become strong and are assumed to produce a meta-stable SUSY violating state\footnote{To ensure this, it may be necessary to introduce quadratic terms in the superpotential by hand \cite{ISS}.  Depending on one's theoretical orientation, one may view these as arising from retro-fitting \cite{DFS} or from Cosmological SUSY Breaking \cite{CSB}.}.

One of the phenomenological virtues of the MSSM is its successful prediction of coupling constant unification.  If we wish to preserve this prediction, to one loop order, then the $G$-charged chiral fields must lie in complete multiplets of the unified group.  Furthermore, there are strong constraints on the gauge group $G$, and the additional matter content, from the requirement that the standard model gauge couplings remain in the perturbative regime all the way up to the GUT scale.  As far as we know, the only phenomenologically viable choice of $G$ which {\it might} satisfy these constraints is $SU(5)$, and one is led to the Pentagon model \cite{pentagon}.  Even in the Pentagon model the dynamics which leads to a phenomenologically viable SUSY violating state is somewhat conjectural.  In all other examples that we have studied, there are dramatic clashes with existing experiments - spontaneous breakdown of charge or color, or unobserved light states.

Recently, a careful two loop study of the standard model running couplings has shown \cite{jj} that the Pentagon is viable only if the scale $\Lambda_5$, and the ISS mass terms are both $>1000$ TeV.  This is incompatible with the original motivation for the Pentagon model, in which it was the low energy implementation of Cosmological SUSY Breaking.  For most readers it will be more significant that the lower bound on the SUSY breaking scale pushes up against the forbidden window of gravitino masses.  A conservative reading of the literature on cosmological gravitino bounds leads one to conclude that $m_{3/2}<30$ eV, corresponding to a bound on the highest SUSY breaking scale of order $\sqrt{6}\times10^2$ TeV.  If we raise the scale high enough to get to the high side of the forbidden window for gravitino masses, then we lose the solution to the SUSY flavor problem.

Yet another problem with the Pentagon model surfaced in a recent paper \cite{tbhh}.  The pseudo Nambu-Goldstone boson of spontaneously broken penta-baryon number, gets its mass from an operator of dimension $7$.  If the scale associated with this irrelevant operator is larger than $\sim 10^{10}$ GeV then the PNGB is copiously produced in stars and leads to unobserved stellar cooling\footnote{It should be noted that if one postulates a scale $\sim10^8-10^{10}$ GeV for the coefficient of the dimension $7$ operator, and also a primordial asymmetry in penta-baryon number, then one can get a unified explanation of the baryon asymmetry of the universe, and the origin of dark matter \cite{bej}.}.

Finally, like most gauge mediated models, the Pentagon model does not have a SUSY WIMP dark matter candidate.  One is forced to invoke either a QCD axion, or the scenario mentioned in the previous footnote.

In this paper we will show that all of these problems can be solved simultaneously if we replace unification in $SU(5)$ or some larger group, with {\it trinification} \cite{trinify}.  We will present an explicit direct mediation model called The Pyramid Scheme, which realizes these ideas.  However, we note that the idea of resolving the Landau pole problem of direct mediation with trinification, may be of more general utility.

\subsection*{Trinification and the Pyramid Scheme}

In $E_6$, one generation of standard model fermions is embedded in the $[27]$ representation.  $E_6$ has an $SU_1(3)\times SU_2(3)\times SU_3(3)\rtimes Z_3$ subgroup, under which
\begin{equation*}
[27] =(3,1,\bar{3}) \oplus (\bar{3},3,1) \oplus (1,\bar{3}, 3),
\end{equation*}
with the three groups and representations permuted by the $Z_3$.  $SU_3(3)$ is identified with color, while the electro-weak $SU(2)$ is the upper Cartesian subgroup of $SU_2(3)$.  Weak hypercharge is a linear combination of the hypercharge generators of the first and second $SU(3)$ factors.  The usual $15$ components of the $[27]$ make up a standard model generation, while the Higgs fields $H_{u,d}$ of the MSSM can be obtained in a variety of ways from $[27]$ and $[\bar{27}]$ representations of $E_6$.

The essential idea of trinification, is that, in order to predict gauge coupling unification, it is sufficient, at one loop, to insist that all extra matter between the weak scale and the unification scale, fall into complete multiplets of $SU(3)^3\rtimes Z_3$, and that there be no strong breaking of this symmetry by Yukawa couplings.  The latter requirement is subsumed under the further demand that {\it all} couplings remain perturbative up to the unification scale, so that one loop renormalization group formulae are a good approximation\footnote{Two loop unification in the MSSM works less well than one loop unification, and is subject to unknown unification scale threshold corrections, so we do not consider two loop unification to be a necessary desideratum of a good model.}.

Although we have described trinification in terms of embedding in an underlying $E_6$, it might also be derived in a simple manner from D-brane constructions in Type II string theory, or related geometric engineering models \cite{dbranetrini}.  This notion makes the {\it Pyramid Scheme}, which we now introduce, particularly natural.

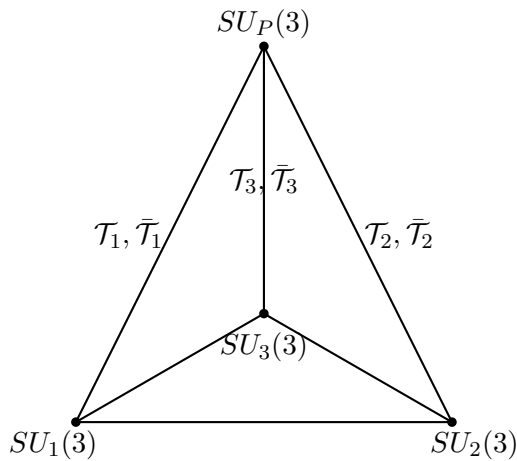
\begin{figure}[t]
\begin{center}
\begin{pspicture}(0,0)(7,7)
\psdot(1,1)
\psdot(6,1)
\psdot(3.5,2.443375673)
\psline(1,1)(6,1)
\psline(6,1)(3.5,2.443375673)
\psline(3.5,2.443375673)(1,1)
\psdot(3.5,6)
\psline(1,1)(3.5,6)
\psline(6,1)(3.5,6)
\psline(3.5,2.443375673)(3.5,6)
\rput(0.7,0.7){{\small $SU_1(3)$}}
\rput(6.3,0.7){{\small $SU_2(3)$}}
\rput(3.5,2.0){{\small $SU_3(3)$}}
\rput(3.5,6.3){{\small $SU_P(3)$}}
\rput(1.7,3.5){{\small ${\cal T}_1,\bar{{\cal T}}_1$}}
\rput(5.3,3.5){{\small ${\cal T}_2,\bar{{\cal T}}_2$}}
\rput(3.5,4.221687836){{\small ${\cal T}_3,\bar{{\cal T}}_3$}}
\end{pspicture}
\caption{Quiver Diagram of the Pyramid Scheme.  Standard Model Particles are in broken multiplets running around the base of the pyramid.}
\end{center}
\label{fig:quiver}
\end{figure}

In the Pyramid Scheme we extend the quivering moose of trinification by a fourth $SU(3)$ group, $SU_P(3)$.  All standard model fields are singlets of the new group, and we add the new representations
\begin{eqnarray*}
{\cal T}_1 +\bar{\cal T}_1 &=& (3,1,1,\bar{3}) + (\bar{3},1,1,3),\\
{\cal T}_2 +\bar{\cal T}_2 &=& (1,3,1,\bar{3}) + (1,\bar{3},1,3),\\
{\cal T}_3 +\bar{\cal T}_3 &=& (1,1,3,\bar{3}) + (1,1,\bar{3},3).
\end{eqnarray*}
We call these new matter fields, {\it trianons}.  Note that only the third trianon carries color.  Thus, the one loop running of the gauge couplings will be like that in a vanilla gauge mediated model with $3$ messengers.  One loop perturbative coupling unification will be preserved.  The quivering moose of this model has the pyramidal shape of figure \ref{fig:quiver}, which accounts for the name.

In a D-brane or geometric engineering construction, trinification corresponds to 3 singular loci (stacks of wrapped D-branes) residing on a set of internal cycles which are permuted by a $Z_3$ isometry of the compact geometry.  We call these the chiral cycles since the chiral fields result from topological intersections of these cycles.  The Pyramid Scheme introduces an extra stack of branes, wrapped on a cycle with the appropriate (non-topological) intersection with each of the chiral cycles.  The trianon mass terms that we introduce below correspond to small deformations of this extra cycle, so that it no longer intersects the chiral cycles.

As in the Pentagon model, we introduce a chiral field $S$, singlet under all gauge groups, with superpotential couplings
\begin{equation*}
W_S = g_{\mu}SH_uH_d+\frac{g_T}{3}S^3+\sum_{i=1}^3y_iS{\cal T}_i\bar{\cal T}_i,
\end{equation*}
where the bilinears in the trianon fields are the unique $SU(3)^4$ invariants.  The $Z_3$ symmetry imposes $y_i=y$, independent of $i$.  Strictly speaking, we do not have to impose this much symmetry on the Yukawa couplings, if they are sufficiently small, because they affect gauge coupling running only at two loops.  The only inviolable symmetry of this low energy Lagrangian is the low energy gauge symmetry $SU(1,2,3)\times SU_P(3)\times Z_R$\footnote{$Z_R$ is the discrete $R$-symmetry required by CSB.  We also use it to forbid unwanted dimension 4 and 5 operators in the MSSM.  We will discuss it in section 2, below.}.  For simplicity however we will assume that the full Pyramid gauge group is broken only by the part of the Lagrangian containing standard model fields, and by the Intriligator-Seiberg-Shih (ISS) \cite{ISS} trianon mass terms.  It is certainly worth exploring more complicated models, in which the gauge symmetry is broken down to the standard model ($\times SU_P(3)$), also in the couplings to $S$.

The singlet $S$ serves several purposes in the model.  Most importantly, the term $|\frac{\partial W}{\partial S}|^2$ ties $SU_L(2)\times U_Y(1)$ breaking to the properties of the meta-stable SUSY violating state of the strong $SU_P(3)$ gauge theory.  This predicts $\tan\beta\sim1$ for the Higgs mixing angle.  Secondly, the VEV of $S$ can give rise to the $\mu$ term of the MSSM, while $F_S$ generates the $B_{\mu}$ term.  We will discuss mechanisms for generating such VEVs below.  We note that the coupling $g_{\mu} $ can ameliorate the little hierarchy problem, but that this might interfere with our desire for a VEV of $S$.

The rest of this paper is organized as follows.  In the next section we find a discrete $R$-symmetry of the Pyramid model, which outlaws all dimension four and five $B$ and $L$ violating couplings, apart from the neutrino seesaw operator.  In section 3 we introduce the ISS mass terms and explore the resulting dynamics of the $SU_P(3)$ gauge theory.  We work in the regime where the mass terms for ${\cal T}_{1,3}$ are above the $SU_P(3)$ confinement scale $\Lambda_3$, while that for ${\cal T}_2$ is close to it.  This produces a non-trivial K\"{a}hler potential for $S$, and reduces the dynamics to the moduli space non-linear $\sigma$ model for the $N_F=N_C=3$ gauge theory with a small mass for the chiral fields.  As in the Pentagon model, we assume a meta-stable SUSY violating state of this system, with VEVs for both the pyrma-baryon and pyrmeson fields constructed from ${\cal T}_2$.  We argue that the extra terms in the potential for $S$, which come from integrating out ${\cal T}_{1,3}$ could lead to a non-zero VEV for this field, if $F_S$ is non-zero.  We also find that the gaugino and squark spectra are ``squeezed" relative to vanilla gauge mediation models \cite{shihetal}, because the colored messengers have a SUSY preserving mass higher than the SUSY breaking scale.  We give rough estimates of superpartner masses in this model.

In section 4 we argue that the pyrma-baryons made from ${\cal T}_{1,3}$ could be dark matter, if they are produced in the late
decay of some other particle with a reheat temperature in the TeV range\footnote{They could also have the requisite density as a
consequence of a primordial asymmetry in one or more of the pyrma-baryon numbers.  However, in this case there would be no annihilation signals.} \cite{bmo}.  The dark matter particles annihilate predominantly to the pseudo Nambu-Goldstone boson (PNGB) of the spontaneously broken pyrma-baryon number, which we call the {\it pyrmion}.  The constituents of the pyrmion do not carry color, and we estimate its mass to be a few MeV, so it can decay only to electrons, positrons, photons and neutrinos.  It is possible that this could account for the various dark matter ``signals" that have accumulated over the past few years, along the lines of \cite{ahftw}.  The mass of the pyrmion is also large enough to avoid constraints from stellar cooling \cite{tbhh}.  Section 5 is devoted to conclusions and to many suggestions for further elaboration of this work.  In Appendix A we sketch the basis for the revised estimate of the relation between the gravitino mass and the cosmological constant, which we used in the computations of superpartner masses in section 3.  In Appendix B we recall, for completeness, the calculation done in \cite{bmo} of the non-thermal relic density of pyrma-baryons and Appendix C shows some computations.

Throughout this paper we will use the abbreviations, c.c. for cosmological constant, SUSY and SUSic for supersymmetry and supersymmetric, CSB for Cosmological SUSY Breaking, PNGB for pseudo Nambu-Goldstone boson, and LEFT for low energy effective field theory.  We will use the phrases {\it heavy trianons} and {\it heavy pyrma-baryons} to refer to states constructed from the fields ${\cal T}_{1,3}$.

\section{Discrete $R$-symmetry: the model}

At low energies, the model is $SU_P(3)\times SU(1,2,3)$ where the SM gauge group can be seen as coming from the subgroup $SU(3)^3\times Z_3\subset E_6$.  In the latter notation, the extra matter fields are
\begin{equation*}
\begin{array}{c|cccc}
 & SU_1(3) & SU_2(3) & SU_3(3) & SU_P(3)\\\hline
{\cal T}_1 & 3 & 1 & 1 & \bar{3}\\
\bar{\cal T}_1 & \bar{3} & 1 & 1 & 3\\
{\cal T}_2 & 1 & 3 & 1 & \bar{3}\\
\bar{\cal T}_2 & 1 & \bar{3} & 1 & 3\\
{\cal T}_3 & 1 & 1 & 3 & \bar{3}\\
\bar{\cal T}_3 & 1 & 1 & \bar{3} & 3\\
S & 1 & 1 & 1 & 1
\end{array}
\end{equation*}
and the model can be represented by the quiver diagram shown in figure \ref{fig:quiver}.  We want to find an approximate discrete $R$-symmetry which is exact in the limit of zero ISS masses.  We will in fact look for a $U_R(1)$, of which we imagine only a discrete $Z_N$ subgroup is fundamental.  A variety of equations below only have to be satisfied modulo $N$.

The superpotential terms we would like to have in our model are
\begin{equation*}
W\supset
S{\cal T}_i\bar{\cal T}_i,\;SH_uH_d,\;H_uQ\bar{U},\;H_dQ\bar{D},\;H_dL\bar{E},\;(LH_u)^2
\end{equation*}
which implies that the $R$-charges satisfy (we denote each $R$-charge by the name of the corresponding field)
\begin{eqnarray*}
{\cal T}_i+\bar{\cal T}_i &=& 2-S\\
H_u &=& 2-H_d-S\\
\bar{U} &=& H_d+S-Q\\
\bar{D} &=& 2-H_d-Q\\
\bar{E} &=& 2-H_d-L
\end{eqnarray*}
plus the extra relation from the neutrino seesaw operator.  The (approximate) $U_R(1)$ anomaly conditions are
\begin{eqnarray*}
SU_P(3)^2U_R(1) &\Rightarrow& 2\cdot3+3({\cal T}_1+\bar{\cal T}_1+{\cal T}_2+\bar{\cal T}_2+{\cal T}_3+\bar{\cal T}_3-6)=3(2-3S)\\
SU_C(3)^2U_R(1) &\Rightarrow& 2\cdot3+6(Q-1)+3(\bar{U}+\bar{D}-2)+3({\cal T}_3+\bar{\cal T}_3-2)=0\\
SU_L(2)^2U_R(1) &\Rightarrow& 2\cdot2+(H_u+H_d-2)+9(Q-1)+3(L-1)\\
 && +3({\cal T}_2+\bar{\cal T}_2-2)=3(3Q+L)-4(S+2)
\end{eqnarray*}
which might allow for an $S^3$ superpotential term if $3S=2\mod N$.

The dangerous higher-dimensional superpotential and K\"{a}hler potential terms can be combined into seven groups (the neutrino seesaw operator is allowed).  Operators in each group have the same $R$-charge (once one takes the $d^2\theta$ for
superpotential terms into account).
\begin{eqnarray*}
G_1=\{LL\bar{E},\;LQ\bar{D},\;SLH_u\} &\Rightarrow& L-H_d\\
G_2=\{LH_u,\;Q\bar{U}\bar{E}H_d,\;\bar{U}\bar{D}^*\bar{E}\} &\Rightarrow& L-H_d-S\\
G_3=\{\bar{U}\bar{U}\bar{D}\} &\Rightarrow& 3Q+H_d-S-2\\
G_4=\{QQQL\} &\Rightarrow& 3Q+L-2\\
G_5=\{QQQH_d,\;QQ\bar{D}^*\} &\Rightarrow& 3Q+H_d-2\\
G_6=\{\bar{U}\bar{U}\bar{D}\bar{E}\} &\Rightarrow& 3Q+L-2S-2\\
G_7=\{LH_uH_dH_u\} &\Rightarrow& L-H_d-2S+2.
\end{eqnarray*}
It is possible to forbid all dangerous terms.  For example, with $N=5$, and $S=4$, $3Q+L=3$, $L=3+H_d$, and any choice of
$H_d$ one finds that all anomaly conditions are satisfied and none of the dangerous terms are allowed.  Notice moreover that the $S^3$ superpotential term and neutrino seesaw operator are allowed by this choice of $R$-charges.  Thus one can engineer a superpotential of the form
\begin{multline*}
W=\sum_{i=1}^3(m_i+y_i S){\cal T}_i\bar{\cal T}_i+g_\mu
SH_uH_d+\frac{g_T}{3}S^3\\
+\lambda_u H_uQ\bar{U}+\lambda_d H_dQ\bar{D}+\lambda_L
H_dL\bar{E} + \frac{\lambda_{\nu}}{M} (L H_u)^2 + W_0
\end{multline*}
where only the ISS masses $m_i$ and $W_0$ break the $R$-symmetry.  Note that in this equation $\lambda_{u,d,\nu}$ are all matrices in generation space.

\section{Breaking $R$-symmetry and SUSY}

We now take into account the dynamical effect of the $R$-symmetry breaking superpotential
\begin{equation*}
\delta W=W_0+m_1{\cal T}_1\bar{{\cal T}}_1+m_2{\cal T}_2\bar{{\cal T}}_2+m_3{\cal T}_3\bar{{\cal T}}_3
\end{equation*}
to the low energy effective Lagrangian.  Using conventional effective field theory philosophy, we could ascribe this by the strategy of {\it retro-fitting} \cite{DFS}.  That is, we imagine that the $R$-symmetry breaking occurs spontaneously, as a consequence of strong dynamics at a scale $\Lambda_R\gg\Lambda_3$ and that the mass terms arise from irrelevant couplings between this sector and the Pyramid model, and have a size $m_i\sim\frac{\Lambda_R^{d_R}}{M^{d_R - 1}}$, where $d_R$ is the dimension of the operator appearing in lowest dimension $R$-conserving coupling of the two sectors.  $M$ could be either the
unification scale or the Planck scale, depending on one's microscopic model for these couplings.  $W_0$ is simply added as a phenomenological fudge to obtain the right value of the cosmological constant.  Apart from the exigencies of phenomenology, there is no requirement in this way of thinking, that the operators to be added create a SUSY violating meta-stable state in the low energy theory.  Indeed, if one adds operators which do create such a state, one must be careful to engineer the model so that these are the dominant effects of the coupling between the two sectors.

The explanation for $\delta W$ on the basis of the hypothesis of CSB has a very different flavor.  Here, the size of the c.c. and the relation $m_{3/2}=10K\Lambda^{1/4}$ \footnote{See Appendix A for an explanation of the new factor of $10$ in this equation.  $K$ is for the moment, a ``parameter of order 1", which cannot be determined from first principles.}, are fundamental inputs of a microscopic theory of quantum de Sitter space.  In order to be consistent with this theory the low energy effective Lagrangian {\it must} have a meta-stable SUSY violating state\footnote{And the Lagrangian must be {\it above the Great Divide} \cite{abj} so that transitions out of this state can be viewed as highly improbable Poincar\'{e} recurrences of a low entropy state in a finite system, rather than as an instability.}.  Furthermore, $\Lambda$ is prescribed by the microscopic theory, and the tuning of $W_0$ simply implements this prescription in the LEFT.

The $SU_P(3)$ gauge theory is IR free with a small $\beta$ function.  Starting from some unification scale boundary condition, the coupling decreases slowly in the IR.  If there were no mass terms $m_i$ it would flow to a free theory and SUSY would be preserved.  This could not be the low energy implementation of CSB.  We must thus introduce mass terms, in order to produce a dynamical meta-stable SUSY violating state with $m_{3/2}=10K\Lambda^{1/4}$.  In order to do this using the known and conjectured dynamics of $N_F\geq N_C$ SUSY QCD, we take two masses $m_{1,3}$ somewhat larger than the third, $m_2$.  The gauge coupling then becomes strong at a confinement scale $\Lambda_3$, and we assume that $m_2$ is small enough to be treated by chiral perturbation theory in the $N_F=N_c=3$ moduli space Lagrangian\footnote{Another possibility is to take $m_3>m_{1,2}$.  The theory then flows close to an interacting superconformal fixed point and for some range of parameters we may find a calculable meta-stable state.  We thank N. Seiberg for explaining this possibility to us.  We leave the exploration of this scenario to future work, but note that the meta-stable state has the approximate $R$-symmetry of ISS vacua, and may be phenomenologically problematic.}.  We must further assume that the unification scale coupling is large enough that $m_3/\Lambda_3$ is not too large.

The latter assumption, and the choice of $m_3$ as one of the large masses, is motivated by phenomenology.  We will see that taking $m_3$ somewhat larger than $\Lambda_3$ solves one of the fine tuning problems of vanilla gauge mediation.  It suppresses the gluino/chargino mass ratio.  If $m_3$ is too large, this suppression produces an unacceptably light gluino.

We can think of the two heavy trianons as analogs of the charmed quark in QCD, while the light one is analogous to the strange quark.  For purposes of assessing the nature of the (meta-stable) ground state, we integrate out the heavy trianons, and treat the LEFT by chiral (moduli space) Lagrangian techniques.

For phenomenological reasons, we will take the two heavy trianons to be ${\cal T}_{1,3}$.  As a consequence the light moduli are color singlets and will give rise to gaugino masses only for the electro-weak gauginos.  The gluino mass will be induced by a SUSY breaking mass for ${\cal T}_3$, and will be suppressed relative to the chargino masses because this field has a relatively large supersymmetric mass term.  This relieves the tension between the experimental lower bound on the chargino mass (which might soon reach $160$ GeV as a consequence of the Tevatron trilepton studies \cite{sunil}), and the large radiative corrections to the Higgs potential coming from heavy gluinos.  There will be a similar suppression of the squark to slepton mass ratio, relative to the predictions of vanilla gauge mediation.

The moduli space of the $SU_P(3)$ gauge theory coupled to ${\cal T}_2$ consists of a $3\times3$ complex matrix pyrmeson field, $M$, transforming in the $[3,\bar{3}]$ of the $SU_L(3)\times SU_R(3)$ chiral flavor group (whose diagonal subgroup contains the action of electro-weak $SU_L(2)\times U_Y(1)$ on the moduli space), and a pair $P,\tilde{P}$ of flavor singlet pyrma-baryon fields which carry opposite values of a new accidental vector-like $U(1)$ quantum number.  These are related by a constraint 
\begin{equation*}
{\rm det}\,M-\Lambda_3P\tilde{P}=\Lambda_3^3,
\end{equation*}
where $\Lambda_3$ is the complex confinement scale of the theory.  The K\"{a}hler potential is of the form
\begin{equation*}
K = |\Lambda_3|^2 h (e_k , x, \tilde{x}),
\end{equation*}
where $h$ is a real permutation invariant function of the variables $e_k$, the eigenvalues of $Y\equiv\frac{M^{\dagger}M}{|\Lambda_3|^2}$\footnote{Equivalently, a function of $w_k\,{\rm tr}\,Y^k$, for $k = 1,2,3$.}, and of
\begin{equation*}
x = \frac{|P|^2}{|\Lambda_3|^2},\;\;\;\;\;\;\tilde{x}=\frac{|\tilde{P}|^2}{|\Lambda_3|^2}.
\end{equation*}

The superpotential in the chiral LEFT is $W=W_0+m_2\Lambda_3\,{\rm tr}\,M $.  The matrix $M$ can be expanded as $M=Z\sqrt{\frac{2}{3}}I+Z_a\lambda^a$, where the $\lambda^a$ are the Gell-Mann matrices.  We will look for $SU(3)$ invariant states, where $Z_a=0$.  The constraints on the moduli space then imply that $\left(\frac{2}{3}\right)^{3/2}Z^3=\Lambda_3P\tilde{P}+\Lambda_3^3$.  The superpotential is proportional to $Z$ and the locus $P\tilde{P}=0$ is supersymmetric.  Any SUSY violating meta-stable state will have a non-zero VEV for the pyrma-baryon fields, which we will assume charge conjugation symmetric $\tilde{P}=P$.  The constraint then allows us to write both the K\"{a}hler potential and superpotential in terms of the unconstrained complex field $Z$.  Our previous remarks about the structure of the K\"{a}hler potential imply that it is a function of $Z^\dagger Z$, and that the effective potential is
\begin{equation*}
K_{Z^\dagger Z}^{-1}|m_2 \Lambda_3|^2.
\end{equation*}
The existence of a SUSY violating minimum is guaranteed if the positive function $K_{Z^\dagger Z}$ has a maximum at some finite $Z$.  Geometrically, we have a non-compact, circularly symmetric $2$-manifold, and we are asking that the length of a tangent vector attains a maximum at some particular radius.  We have not been able to find arguments for or against the existence of such a maximum, so we will simply explore the phenomenology of the model, under the assumption that the maximum exists.

It should be noted that we have made several assumptions about the symmetry of the ISS mass terms and of the pyrmeson VEV, which are not required by either fundamental principles or phenomenology.  All we are required to preserve in the LEFT is the standard model gauge group, and enough of the trinification structure to guarantee gauge coupling unification.  Thus, there is actually a rich class of pyramid schemes to explore in search of a meta-stable state.  We only treat the most symmetric of them in this paper.

Given our assumptions, the Pyramid model has two kinds of messengers of gauge mediation, the moduli of the $N_F=N_C=3$ theory, and the heavy trianons.  The scalar fields $Z_a$ will get SUSY violating masses of order $m_2$, which, apart from $SU(3)$ symmetry, are completely unconstrained and unconnected with the masses of their fermionic partners.  Therefore we will obtain one loop masses for the $SU_L(2)\times U_Y(1)$ gauginos, of order
\begin{equation*}
m_{1/2}^i=3X_i\frac{\alpha_i}{4\pi}m_2.
\end{equation*}
The $X_i$ are ``order one" numbers, which cannot be calculated without complete knowledge of the K\"{a}hler potential, and the factor of $3$ is the dimension of the fundamental representation of $SU_P(3)$.  The LEFT of the $Z$ fields has quartic scalar couplings of order $(m_2/\Lambda_3)^2$, so we have a consistent low energy expansion only for
\begin{equation*}
m_2<\sqrt{4\pi}\Lambda_3.
\end{equation*}
Combining the estimate above with the gravitino mass formula
\begin{equation*}
m_{3/2}=X_gm_2\Lambda_3/m_P=10K\Lambda^{1/4},
\end{equation*}
gives several competing equalities and inequalities.  Here $X_g$ is a constant which must be calculated from the strongly interacting $SU_P(3)$ gauge theory, while $K$ is a constant of order $1$, which must be calculated from the as yet incomplete quantum theory of de Sitter space.

Plausible model independent extensions of the Tevatron trilepton analysis might eventually bound the charged gaugino mass term from below by $160$ GeV, which requires
\begin{equation*}
19.7<X_2\frac{m_2}{{\rm TeV}}.
\end{equation*}
To get an idea of how these bounds work, assume that $m_2=1.7\Lambda_3$ so that the moduli space Lagrangian is fairly strongly coupled, with a ``fine structure constant" of order $1/4$.  Then $m_2 = 14.9\sqrt{K/X_g}$ TeV and we must have $X_2>1.32\sqrt{X_g/K}$ in order to satisfy the chargino mass bound.  Setting the square root to $\sqrt{3}$ we obtain $m_2=8.6$ TeV and $\Lambda_3=5.1$ TeV.

The heavy trianons, ${\cal T}_{1,3}$ will also have SUSY violating masses, because of their $SU_P(3)$ couplings to the low energy theory.  In particular, since ${\cal T}_3$ carries color, we will get squark and gluino masses.  In the limit where the SUSic masses of the heavy trianons are $\gg\Lambda_3$, we could calculate the resulting gluino masses by integrating the heavy
trianons out to create effective couplings of the form {\it e.g.}
\begin{equation*}
\int d^2\theta\;(W_{\alpha}^{(3)})^2f(P/m_3,\tilde{P}/m_3,M/m_3).
\end{equation*}
The F-terms of the light fields would then generate small gluino masses.  Symmetries imply that the leading operators are fairly high-dimensional.  However, there is no reason to suppose that $m_{1,3}\gg\Lambda_3$.  For example, in ordinary QCD, an hypothetical quark with mass of order the rho meson mass, would not be treated by chiral perturbation theory, but neither would it make sense to estimate its effects via the operator product expansion.  Thus, we predict a gluino/chargino mass ratio which is definitely smaller than the vanilla gauge mediation result $\alpha_3/\alpha_2$, and depends sensitively on $m_3/\sqrt{m_2\Lambda_3}$ as that variable becomes large.  There will be a similar suppression of the squark to slepton mass ratios. A factor of $2$ in $m_3/m_2$ could easily bring the gluino and squark mass predictions down to the range where they are consistent with experimental lower bounds but do not give large contributions to the Higgs potential.  The mass $m_1$ is not constrained by this analysis.

In connection with the sfermions, we remark that our messenger mass spectrum does not satisfy any super-trace sum rules.  This has two consequences.  First, it implies that sfermion masses will be logarithmically divergent in the LEFT, which is non-renormalizable\footnote{Note that the arguments for the cancelation of this divergence in \cite{ggm} depended on having conformally invariant short distance behavior.  This will be true in the full Pyramid Scheme, but is not true at the scale of the low energy chiral Lagrangian.} \cite{tbjj}.  It also implies that there are no {\it a priori} arguments that sfermion squared masses are positive, since those arguments depend on the same sum rules.  We have to worry about the possibility of charge, color or lepton number breaking minima for the MSSM sfermions, but will not do so in this paper.

We will of course assume that no disasters happen.  The logarithmic enhancement of the right handed slepton mass, relative to that of the bino, suggests that the bino will be the NLSP, which would imply that LHC will see events with hard $X+l^+l^-\gamma\gamma$, plus missing transverse energy.  The origin of these events is the decay of a slepton to the bino and a hard lepton, followed by bino decay to a photon and a longitudinally polarized gravitino.  Depending on the structure of the SUSY cascade, we will have other particles, denoted by $X$ in the final state.  At LHC strong production cross sections for sparticles dominate, so we might expect $X$ to include at least a dijet.  If the cascade passes through the relatively light chargino then there will be $W$ bosons in $X$, coming from the decay of the chargino into $W$ plus neutralino.  The leptons might not even be hard.  So the general characterization of final states for a bino NLSP is $X$ plus two hard photons plus missing transverse energy, where $X$ depends on the nature of the SUSY cascade.

The ratio $m_{1/2}^1/m_{1/2}^2$ is given by
\begin{equation*}
\frac{m_{1/2}^1}{m_{1/2}^2}=\frac{X_1\alpha_1}{X_2\alpha_2}=0.5\frac{X_1}{X_2}.
\end{equation*}
It's clear that we can only predict these masses up to a factor of a few.  Unfortunately, the unknown strong interaction factors might well affect the phenomenological signals of our model.

The ratio of the right handed slepton mass to that of the bino is $f=Y\ln^{1/2}(\sqrt{4\pi}m_2/\Lambda_3)$.  $Y$ is another unknown strong interaction factor, and we have used the usual naive dimensional analysis estimate of the cutoff for the moduli space LEFT.  If we take $Y=\sqrt{3}$  and $m_2=1.7 \Lambda_3$, then $f=2.3$, while for $Y=1$ and $m_2=\Lambda_3$ we have $f=1.12$.  It seems likely that the bino will be the NLSP in the Pyramid model.  For a $50$ GeV bino we need $f\gtrsim2$ in order to satisfy the experimental bound on the right handed slepton mass.

\section{The Higgs sector and $SU_L(2)\times U_Y(1)$ breaking}

In the approximation that the two heavy trianon masses are $\gg\Lambda_3$, integrating out the trianons and $SU_P(3)$ gauge bosons leads to two distinct contributions to the effective action for the Higgs sector of the NMSSM.  The heavy trianon couplings to $S$ give us a non-trivial effective potential for $S$.  In the Coleman-Weinberg (CW) approximation it has the form
\begin{equation*}
\sum_{i=1,3}|m_F^i|^4 f(u_i).
\end{equation*}
Here $m_F^i=m_i+y_iS$ and
\begin{equation*}
u_i\equiv\frac{|F_S|^2}{|m_F^i|^4}.
\end{equation*}
This expression is valid if the $y_i$ are perturbative and $u_i<1$.  We have
\begin{equation*}
f(u)=au-\sum_{n=0}^{\infty}\frac{u^{n+2}}{(n+1)(n+2)(2n+3)}.
\end{equation*}
The linear term comes from the logarithmically divergent one loop wave function renormalization for $S$.  The rest of the potential is a negative, monotonically decreasing convex function of $u_i$, which becomes complex at $u_i=1$.  This change of behavior represents the breakdown of effective field theory when the masses of scalar components of the heavy trianon fields become smaller than other scales in the theory, like $\Lambda_3$ and $m_2$.  Calculation of the potential in this regime is more complicated.  Note that when $F_S\neq0$, the CW potential monotonically decreases as $m_F^i$ are lowered.  Thus, these contributions tend to make the $S$ VEV non-zero when $F_S\neq0$.  This tendency competes against the contributions to the potential from Higgs F-terms, which are proportional to $|g_{\mu}|^2$.

The second contribution to the Higgs potential is the non-zero pyrmeson VEV $({\cal T}_2\bar{{\cal T}}_2 )^i_j\sim\Lambda_3Z\delta^i_j$.  The resulting Higgs potential, including standard model D-terms has an $SU_L(2)\times U_Y(1)$ breaking minimum with
\begin{eqnarray*}
g_{\mu}H_uH_d &=& \sqrt{6}y_2\Lambda_3Z,\\
\tan\beta &=& 1,
\end{eqnarray*}
and
\begin{equation*}
S=F_S=0.
\end{equation*}
This minimum breaks SUSY and $R$-symmetry because of the VEVs of $Z$ and $F_Z$.  Given our estimate $\Lambda_3\sim5$ TeV, we need $y_2\sim0.01$, a perfectly reasonable value for a Yukawa coupling.  We assume that all Yukawa hierarchies in the model are explained in terms of unification scale physics, a point of view motivated by the strict bounds on flavor changing processes.

When we include quantum corrections to the potential from loops of high scale $SU_P(3)$ gauge bosons, we obtain couplings between $S$ and $Z$.  We have not calculated these, but if they have the effect of forcing $F_S\neq0$, due to a coupling to $F_Z$, then the VEV of $S$ is likely to shift as well.  Thus it is at least plausible that we obtain MSSM $\mu$ and $B_{\mu}$ terms of the right order of magnitude.

The lower bound on the gluino mass implies that the approximation $m_3\gg\Lambda_3$ is unlikely to be valid.  Rather, it is likely that $m_3$ should be thought of as the moral equivalent of a quark mass of order $800$ MeV in QCD: too large to be treated by chiral perturbation theory, but too small to integrate out above the confinement scale.  In other words, the CW approximation we discussed above is probably inadequate, if the model is to produce an acceptably large gluino mass.  The generation of effective $\mu$ and $B_{\mu}$ terms is thus mixed up with the strong $SU_P(3)$ gauge dynamics.

To summarize: we have given plausibility arguments that, in an appropriate range of the parameters $m_i$, the Pyramid Scheme has a SUSY violating, $R$-symmetry violating meta-stable minimum with a non-zero value for $S$.  It can give rise to a reasonable supersymmetric phenomenology, but detailed calculation of the superpartner spectrum is not possible at this juncture, though it seems likely that a neutralino is the NLSP.

We end this section with a discussion of the tuning of parameters in our model, and its interpretation.  Although we do not have a precise calculation of superpartner masses, it seems possible that the Pyramid Scheme does not suffer from a {\it little hierarchy problem}.  It incorporates the NMSSM and the Yukawa coupling $g_{\mu}SH_uH_d$ can evade the usual bounds on the lightest Higgs mass, even for $\tan\beta\sim1$.  We have presented a mechanism that might generate a VEV for $S$, and thus an effective $\mu$ term, despite the substantial size for $g_{\mu}$.  The F-terms of both $S$ and the light pyrmeson can provide a $B_{\mu}$ term of the requisite order of magnitude.

Our required pattern of two trianon masses slightly above $\Lambda_3$, with the third in the range of validity of chiral perturbation theory may seem artificial, but in the CSB interpretation of the Pyramid Scheme it is in fact required in order to reproduce the meta-stable state implied by the underlying (but still partly hypothetical) quantum theory of dS space.  Perhaps
retro-fitters of the Pyramid Scheme would be more hard pressed to justify precisely this pattern of masses, but it is surely no more bizarre than the actual pattern of quark and lepton masses in the standard model.

Finally we note that, as a flat space field theory, the Pyramid Scheme certainly has supersymmetric AdS minima in its effective gravitational action.  Given our instructions to tune $W_0$ so that the cosmological constant in the meta-stable state is almost zero, these states could at best correspond to AdS theories of quantum gravity (superconformal $2+1$ dimensional field theories) with cosmological constant of order $-|m_2 \Lambda_3|^2$.  They have nothing to do with the evolution of our meta-stable state, and belong to a different quantum theory of gravity, with a different Hamiltonian, if they exist at all.  The actual ``decay" of the meta-stable de Sitter state proceeds to a Big Crunch space-time in which the low energy effective description breaks down.  Two features of this breakdown are worthy of note.  First of all, high energy degrees of freedom of the field theory are excited.  In particular, even in the moduli space approximation, the fields do not remain in the vicinity of the negative c.c. minimum, but instead explore the entire potential.  Secondly the covariant entropy bound restricts the entropy observable by any observer in the crunching region to be less than $\sim\frac{M_P^2}{m_2\Lambda_3}$.

In \cite{abj} it was shown that the space of potentials exhibiting ``de Sitter decay" is divided into two classes.  In the first class, {\it above the Great Divide}, the decay probability behaves like $e^{-\pi(RM_P)^2}$ for large de Sitter radius.  These transitions look more like Poincar\'{e} recurrences, temporary sojourns in low entropy states of a finite system, than like true decays.  This is consistent with the hypothesis of Fischler and one of the present authors (TB) that a stable dS space has a finite number of states.  It is also consistent with the low entropy implied for the crunching region by the covariant entropy bound.  Thus, within a class of potentials for a meta-stable dS minimum in field theory, the semi-classical dynamics is consistent with the idea of a stable quantum dS space with a finite number of states.  The instability of the semi-classical theory is viewed as a Poincar\'{e} recurrence.  The parameters of the Pyramid Scheme must be chosen to lie in the regime above the Great Divide, where this analysis is applicable.  There is no corresponding {\it a priori} restriction, if we view the Pyramid Scheme's $R$-violating parameters as arising from retro-fitting in the conventional view of effective field theory.

\section{A Pyramid Scheme for cosmology}

Models of gauge mediated SUSY breaking do not have a standard WIMP dark matter candidate.  Even in the absence of $R$-parity violation, the LSP is the gravitino, which is very light.  When one imposes the further restriction of consistency with CSB, the gravitino mass is about $10^{-2}$ eV.  In \cite{bmo}, Banks and collaborators proposed that baryon-like states of the hidden sector could play the role of cold dark matter.  For reheat temperatures above the confinement scale of the hidden sector, this was only possible if there was a primordial asymmetry in the hidden sector baryon density.

The discovery of the ISS \cite{ISS} meta-stable vacua did not fit in with this idea, because in these states SUSY breaking is correlated with spontaneous breakdown of the hidden sector baryon number\footnote{This correlation persists for the $N_f=N_C$
models, which might have vacua breaking the discrete $R$-symmetry of the ISS states.}.  In \cite{bej}, with another set of collaborators, Banks proposed that the PNGB of the spontaneously broken hidden sector baryon number could be the dark matter.  This was only possible if there was a primordial asymmetry in this quantum number.  Such an asymmetry would automatically generate an ordinary baryon asymmetry, through the mechanism of spontaneous baryogenesis \cite{ck}, because of the effective coupling of the hidden sector and ordinary baryon number currents, due to gluon exchange.  If one bounds the hidden sector asymmetry by insisting that the ordinary baryon asymmetry is no bigger than what is observed, then the dark matter density is also bounded, though the bound is model dependent, and depends on the scale at which hidden sector baryon number is broken.  In the Pentagon model, one had to assume the scale associated with the leading penta-baryon number violating operator was between $10^{8}-10^{10}$ GeV, in order to explain the observed dark matter density.

A related astrophysical issue with the PNGB was pointed out in \cite{tbhh}.  Rather general arguments show that the effective Yukawa coupling of the PNGB to electrons, violates bounds coming from stellar cooling rates.  To avoid this, one must raise the mass of the PNGB to about an MeV, so that it cannot be produced in ordinary stars.  In the Pentagon model this again required the scale associated with the leading symmetry violating operator to be in the $10^{8}-10^{10}$ GeV range.

The Pyramid Scheme throws a new light on all of these questions.  It has three accidental baryon number like symmetries, corresponding to the three types of trianon.  Call the corresponding conserved charges ${\cal B}_i$.  The dynamics of $SU_P(3)$ spontaneously breaks ${\cal B}_2$, but the other two are preserved.  The lightest particles carrying ${\cal B}_{1,3}$ are standard model singlets, and thus potential dark matter candidates.  According to \cite{bmo} there is a small window of low reheat temperatures, below the confinement scale of $SU_P(3)$ in which non-thermal production of these particles could account for the observed dark matter density\footnote{We recapitulate this analysis in Appendix B.}.  Alternatively, a primordial asymmetry in any of these quantum numbers could be invoked to explain dark matter in a cosmological model with high reheat temperature.  One would have to correlate this with the ordinary baryon asymmetry, as in \cite{bej}, a constraint which was missed in \cite{bmo}.  Whether or not there is a PNGB, a primordial asymmetry in some quantum number implies a cosmological expectation value for the associated charge density.  The ${\cal B}_i$ currents are all coupled to the ordinary baryon number current via exchange of standard model gauge bosons, and, in combination with electro-weak baryon number violation, the asymmetries in pyrma-baryon numbers can drive spontaneous baryogenesis.

The Pyramid Scheme thus provides us with a wealth of possibilities for explaining both the dark matter in the universe and the asymmetry in ordinary baryon number.  In this paper we will only explore one of these directions.  We assume that only negligible primordial asymmetries in any of these quantum numbers were generated in the very early universe, and assume a low reheat temperature, so that particles carrying ${\cal B}_{1}$ and/or ${\cal B}_3$, can be the dark matter.

These particles have QCD like strong interactions, with confinement scale $\Lambda_3$.  Their annihilation cross section is energy independent and of order $\Lambda_3^{-2}$.  Probably the best model for their cosmological behavior is the soliton picture of \cite{grikam}.  By analogy with baryon anti-baryon annihilation in QCD, and more generally with soliton anti-soliton annihilation, we expect the typical final state of the annihilation process to be a state of pyrmions (the PNGB of spontaneously broken ${\cal B}_2$) with high multiplicity.  This is quite interesting, because the pyrmions are very light (we will estimate their mass below, in the MeV range), and their constituents do not carry color.  As a consequence, the pyrmion decay into standard model particles will primarily produce electron positron pairs, photons and neutrinos.

One is tempted to try to associate the behavior of our hypothetical dark matter candidate, with some of the ambiguous signals for dark matter that have accumulated in recent years \cite{ambisigs}.  In \cite{ahftw} it was emphasized that this data can only be interpreted in terms of a dark matter candidate which decays primarily to leptons, and the authors constructed an ingenious set of models to implement this constraint.  Our suggestion is, quite frankly, modeled on theirs, but fits more organically into the framework of gauge mediated SUSY breaking.  We will only sketch the outlines of it here, since much more work is needed to see whether it is viable.  The Pyramid model in fact predicts a zero temperature cross section for dark matter annihilation which is just what is needed to explain the ATIC, PAMELA and PPB-BETS data.  The dimensional analysis/soliton estimate is an energy independent cross section
\begin{equation*}
\sigma_0=\frac{A}{\Lambda_3^2}.
\end{equation*}
Recall that $\Lambda_3$ was constrained strongly by the twin requirements of an experimentally acceptable chargino mass and a gravitino mass obeying the CSB formula.  A typical value obeying the bounds was $\Lambda_3\sim5$ TeV.

The interpretation of the ATIC, PAMELA, PPB-BETS and WMAP haze data in terms of dark matter annihilation requires a low energy cross section
\begin{equation*}
\sigma_0^{exp}\sim0.1\;({\rm TeV})^{-2} .
\end{equation*}
Thus $A\sim 2.5$ would seem to fit the data.  We will see below that the multiplicity of $e^+e^-$ pairs per dark matter annihilation is likely to be large, so that an even smaller cross section for dark matter annihilation is actually called for.  This would require $A\sim0.2$ for the multiplicity we estimate below.  Our point here is not to make precise fits, but rather to show that the Pyramid Scheme is in the right ballpark to explain the observational evidence for a lepton anti-lepton excess in the galaxy.

Fans of thermal WIMP dark matter will be curious to understand how such a large cross section could be compatible with the correct relic dark matter density.  For completeness, we recapitulate the non-thermal dark matter production calculation of \cite{bmo} in the Appendix.  The answer depends on the last reheat temperature of the universe, which must satisfy
\begin{equation*}
\Lambda_3>T_{RH}>0.1m_{\cal B}.
\end{equation*}
It is easy to imagine getting such a low reheat temperature from the decay of a relic scalar, like the supersymmetric partner of the QCD axion \cite{bdg}.

With a low reheat temperature, we must look for a method of creating the baryon asymmetry of the universe which is efficient at low energy.  Affleck-Dine baryogenesis is always an option \cite{AD}, but the Pyramid model has the possibility of creating the asymmetry via spontaneous baryogenesis \cite{ck} at the electroweak phase transition \cite{bej}.  That is, a primordial asymmetry in any of the pyrma-baryon numbers acts, because of couplings $\alpha_3^2J_{\mu}^{PB}B^{\mu}/\Lambda_3^2$ induced by gluon exchange, as a chemical potential for ordinary baryon number.  This biases electro-weak baryon number violation, which is in equilibrium above the electro-weak phase transition.  The asymmetry is frozen in at $T\sim100\;{\rm GeV}\ll T_{RH}$.

In addition to this, the most suggestive feature in the data is the cut-off on the electron-positron spectrum seen by the ATIC and PPB-BETS detectors \cite{atic}.  In \cite{ahftw} this was interpreted as showing us the mass of the dark matter particle, and gave rise to an estimate $\sim600-800$ GeV.  Our dark matter candidate is $40-60$ times as heavy.

Our proposed explanation for this discrepancy, centers around the strong $SU_P(3)$ interactions of our dark matter candidate, and the existence of the pyrmion PNGB.  Proton anti-proton annihilation at rest, which should be a reasonable analog of heavy pyrma-baryon annihilation in the contemporary universe, produces final states consisting predominantly of pions.  The mean number of pions is $5$, with variance $1$.  Correspondingly, the single pion inclusive momentum distribution is peaked at $0.2$ GeV, roughly $1/5$ of the proton mass.  The experimental peak is pronounced, but reasonably broad.  The distribution has dropped by a factor of $10$ at $0.8$ GeV.  Lu and Amado \cite{luam} have reproduced many of the features of the annihilation data in terms of a soliton model, in which the $p\bar{p}$ initial state is modeled as a zero baryon number lump of pion field in a Skyrme-like model.  Their model gives a peak that is somewhat more narrow than the data.

In a soliton model, the initial state of light mesons after heavy pyrma-baryon annihilation will be a coherent state of the field.  The probability of having $N$ particles in such a state is proportional to the square of the average field strength and the variance is of order $\sqrt{N}$.  In a soliton model of a QCD like theory, the average momentum per particle is strongly suppressed for $|p|>\Lambda_3$, but would otherwise be randomly distributed.  Our dark matter candidate would be a pyrma-baryon consisting of three heavy trianons and would have a mass of order $3m_{1,3}$.  Given our estimates this is roughly $30-40$ TeV.  In the Pyramid Scheme, the final state will consist primarily of pyrmions, which are effectively massless and will have a typical momentum $<\Lambda_3\sim 5$ TeV.  Some of these will be primaries and the rest secondary products of the decays of heavier pyrmesons.  Thus, we may expect the pyrmion multiplicity to be very large and the energy to be thermalized by strong final state interactions.

The single particle momentum distribution of $N$ body massless phase space for annihilation of a particle anti-particle pair with total mass $2M$ is peaked at $|p|=2M/N$ and is a Gaussian of the form
\begin{equation*}
P \propto e^{- a x^2},
\end{equation*}
in the rescaled momentum $x$, around the maximum, with $a=N^2$ for large $N$\footnote{We thank H. Haber for these
results.}.  If we take the estimate $\Lambda_3\sim 5$ TeV from our discussion of superpartner masses, and $m_{1,3}\sim12$ TeV to
assure the massive trianons are outside the range of chiral perturbation theory, then $2M\sim 72$ TeV.  This would give a distribution centered at $800$ GeV, with an extremely narrow width, for $N\sim90$, which is $\sim18$ times the pion multiplicity from proton anti-proton annihilation.  We would interpret the actual distributions seen in the balloon experiments as a broadening of this peak toward the low momentum side by the effects of propagation of electrons and positrons through the galactic medium.  The high side of the experimental peak should be identified with the position of the narrow peak in the primordial distribution.

The underlying $SU_P(3)$ gauge theory is supersymmetric, and has more degrees of freedom than QCD, all of which can decay or annihilate to the pyrmion.  Furthermore, in a soliton model of the annihilation process the probability of a single particle with momentum $>\Lambda_3$ is exponentially suppressed since the particles come from a smooth coherent state.  Thus one would guess that the dynamics of the annihilation process forces a minimum of 10 pyrmions to be produced.  Furthermore, since many of the final state pyrmions will be produced in secondary decays of heavier pyrmesons, the multiplicity is almost certainly higher than 10, since the dynamical momentum cutoff applies to the primaries.  In other words, the high multiplicity required to fit the data on balloon experiments does not seem out of the question.  Obviously, much more work on the dynamics of this strongly coupled annihilation process, as well as a complete model of galactic propagation, will be necessary in order to render a complete verdict on our model of the experiments.

Thus, {\it very roughly} we can produce a spectrum of electrons and positrons consistent with the ATIC, PAMELA and PPB-BETS observations from a heavy pyrma-baryon dark matter candidate decaying into $\sim90$ pyrmions, which themselves decay to $e^+e^-$ pairs.  To be a good candidate for dark matter the pyrma-baryon abundance of the universe must be non-thermal \cite{bmo}, and could come from a late decaying scalar with a reheat temperature in the TeV range.  The details of this, including the relation between the reheat temperature, the low energy annihilation cross section, and the relic abundance, can be found in Appendix B.  The low reheat temperature requires us to invent a sub-TeV mechanism for baryogenesis, and the most attractive candidate is spontaneous baryogenesis at the electro-weak phase transition, driven by a primordial asymmetry in one of the pyrma-baryon numbers \cite{bej}.  Obviously a lot more work is needed to make these remarks into a robust theory, explaining the data on dark matter.

We also note that, should the current observational indications of dark matter annihilation signals prove to be explained by astrophysics \cite{profumo}, the Pyramid Scheme has dark matter scenarios in which there are no annihilation signals.  This would be the case if the dark matter were interpreted as a pyrma-baryon excess, as in \cite{bmo}.  The required primordial asymmetry is roughly $\epsilon_{PB}=\frac{T_{eq}}{m_{PB}}\sim 10^{-12}\frac{\rm TeV}{m_{PB}}$.  This is too small to give rise to an adequate asymmetry in baryon number via spontaneous baryogenesis \cite{ck,bej}.  We could invoke an asymmetry of the spontaneously broken ${\cal B}_2$ quantum number to give spontaneous baryogenesis, but would then have to explain why the inflaton preferred to decay mostly into ${\cal T}_2$ rather than the other trianons.  The Pyramid Scheme can accommodate a wide variety of cosmological scenarios.  We hope to explore some of them in future work.

\subsection{The mass of the pyrmion}

To calculate the mass of the pyrmion, we must understand the way in which the pyrma-baryon number ${\cal B}_2$ is explicitly broken.  The operators
\begin{equation*}
B_2={\rm det}\;{\cal T}_2,\;\;\;\;\;\;\bar{B}_2={\rm det}\;\bar{{\cal T}}_2,
\end{equation*}
are invariant under $SU_P(3)$ and the standard model gauge group, and have discrete $R$-charges satisfying
\begin{equation*}
B_2+\bar{B}_2=3(2-S)\mod N.
\end{equation*}
Recall that $N\geq5$.  We use the freedom to choose the individual pyrma-baryon and anti-pyrma-baryon $R$-charges to impose 
\begin{equation*}
B_2=2-S,\;\;\;\;\;\;\bar{B}_2=4-2S\mod N.
\end{equation*}
In that case, the dimension $5$ operator $\int d^2\theta\,SB_2/M_U$, is the leading ${\cal B}_2$ violating operator, which is invariant under all the symmetries of the model.  The pyrmion mass will then be of order
\begin{equation*}
m_b\sim\frac{\Lambda_3^{3/2}}{{M_U}^{1/2}}\sim2.5\;{\rm MeV}.
\end{equation*}
We have used the estimate $\Lambda_3\sim5$ TeV from our discussion of superpartner masses.  Thus, the pyrmion can decay only into electrons, positrons, photons, neutrinos and gravitinos.

Note that this estimate also resolves the problem of pyrmion production in stars \cite{tbhh}, which could lead to cooling faster
than what is observed.  An MeV scale pyrmion could at best be produced in supernova explosions.

\section{Conclusions}

We have sketched a new Pyramid Scheme for direct mediation of SUSY breaking.  It is based on the same fundamental dynamical assumption as the Pentagon model: the existence of a SUSY and $R$-breaking meta-stable state of $N_F=N_C$ SUSY QCD, but it has the following advantages:

\begin{itemize}

\item{It is based on trinification rather than unification in a simple group, and as a consequence predicts completely perturbative coupling unification, with no Landau poles.  The full model can be associated with a simple quiver/moose diagram, which should make its implementation in string theory straightforward.}

\item{There exist two unbroken baryon-number like symmetries in the hidden sector, which enable us to construct a number of models of dark matter, along the lines of \cite{bmo}.  In this paper we concentrated on a model in which the dark matter is produced non-thermally, but without a pyrma-baryon asymmetry, in order to be able to model dark matter annihilation signals.  Given estimates of the confinement scale of the $SU_P(3)$ gauge group from super-partner masses, the model produces annihilation cross sections of (roughly) the right order of magnitude to explain ATIC, PAMELA and PPB-BETS, and the dark matter annihilates predominantly into pyrmions, the PNGB of the spontaneously broken pyrma-baryon number.  The latter particle has a mass in the MeV range and decays only into light leptons and photons.  We argued that a model of the annihilation process with a high pyrmion multiplicity $\sim100$ in the final state could reproduce the bumps in the ATIC and PPB-BETS data.}

\item{The dark matter scenario requires us to invoke a late decaying particle which reheats the universe to $\sim1$ TeV, which implies that we must supplement it with a low scale model for baryogenesis.  The most economical scheme would be to postulate a primordial asymmetry in one of the pyrma-baryon numbers (not the one associated with the dark matter candidate).  Standard model gauge boson exchange produces current-current couplings between the pyrma-baryon currents and ordinary baryon number, so that a pyrma-baryon asymmetry drives spontaneous baryogenesis \cite{ck} at the electro-weak phase transition.  Affleck-Dine baryogenesis is another reasonable candidate mechanism.}

\item{The pyrmion mass estimate makes it too heavy to be produced in ordinary stars, avoiding the strong constraints of \cite{tbhh} on models of meta-stable SUSY breaking that rely on the dynamics of $N_F\geq N_C$ SUSY QCD.}

\item{The Pyramid Scheme has three pairs of chiral fields, the trianons ${\cal T}_{1,2,3}$ and $\bar{\cal T}_{1,2,3}$, which are charged under the standard model gauge group.  Only one carries color.  In order to generate meta-stable SUSY breaking, the masses of two of the trianons must be too large to be treated by chiral perturbation theory.  If one of these heavy trianons is the colorful one, then the gluino mass is naturally suppressed relative to that of the charginos, and squark masses suppressed relative to those of leptons.  This removes the fine tuning problem of the vanilla gauge mediated spectrum.  We note that the gluino mass goes down rapidly with the mass of the heavy colored trianon, so the latter probably cannot be so large as to be safely integrated out above the confinement scale $\Lambda_3$.}

\end{itemize}

We want to emphasize that our estimates of the properties of the Pyramid Scheme are rather rough and preliminary.  In particular, the discussion of dark matter needs a lot of work before one can make a reliable claim that it accounts for any existing dark matter data.  Furthermore, many of the important dynamical questions in the model, such as the existence of the SUSY and $R$-violating vacuum state, and the generation of appropriate $\mu$ and $B_\mu$ terms, depend on (currently) incalculable strong $SU_P(3)$ dynamics.  The Pyramid Scheme has plausibility, but is not yet a fully going concern.  Investors are warned that past performance is no guarantee of future returns.

\appendix

\section{Cosmological SUSY breaking}

In this appendix we explain the extra factor of $10$, which appeared in our estimate of $m_{3/2}$ according to the hypothesis of CSB.  The variables in the holographic theory of dS space \cite{bfm,bz} are $N\times N+1$ matrices which are also spinors in $7$ compactified dimensions.  We denote them by 
\begin{equation*}
(\psi^P )_i^A,\;\;\;\;\;\;([\psi^{\dagger}]^Q )_B^j.
\end{equation*}
Their quantum algebra is
\begin{equation*}
\bigl{[}(\psi^P )_i^A, ([\psi^{\dagger}]^Q )_B^j\bigr{]}_+=\delta_i^j\delta_B^AM^{PQ}.
\end{equation*}
$P,Q$ are compact dimension spinor indices, and $M^{PQ}$ are ``sums of wrapped brane charges".  Their closed super-algebra with the $\psi$ variables defines the compactification.  We call it the quantum algebra on a single pixel of the holographic screen of dS space, or the {\it pixel algebra} for short.  The holographic principle requires that the pixel superalgebra, for fixed values of $i,j,A,B$ has a finite dimensional unitary representation.  If $D_P$ is the dimension of the pixel algebra representation then ${\rm ln}D_P$ is the entropy per pixel.  The total entropy of dS space, $\pi(RM_P)^2$ is then given by 
\begin{equation*}
\pi(RM_P)^2=N(N+1){\rm ln}D_P.
\end{equation*}
In previous work, $D_P$ was set equal to $2$ because the compactified dimensions were ignored.

We note in passing that this formalism implies that, in a finite radius dS space, compactified dimensions have no moduli.  The finite dimensional algebras and representations are subject to the constraint that, as $N\rightarrow\infty$ we must obtain (super)-gravitons in the spectrum, following the outline in \cite{bfm}\footnote{In fact, in this paper, it was impossible to obtain gravitons (only massless chiral multiplets), because there were no compactified dimensions.}.  The classification of such algebras has not yet been attempted, but they must be discrete.

A hint at what is required comes from noting that Calabi-Yau manifolds are symplectic and compact, so that geometric quantization gives a(n ambiguous) map from their function algebras to finite dimensional matrix algebras.  This can be easily extended to seven manifolds which are Calabi-Yau bundles over an interval (Horava-Witten compactifications) or circle bundles over a $CY_3$.  The variables of the holographic theory will live in modules over these finite dimensional algebras.  From these correspondences one can see that $D_P$ will be related exponentially to the volume of the internal space (in Planck units)\footnote{This is just the statement that entropy is volume extensive in the internal dimensions.  The holographic reduction is just a feature of the non-compact dimensions.}.

What is ${\rm ln }D_P$ in the real world ? In Kaluza-Klein compactification, the volume of the internal space in higher dimensional Planck units (denoted by $M_{Pl}$), is related to the four dimensional reduced Planck scale $m_4$ by
\begin{equation*}
(V/V_{Pl})=(m_4/M_{Pl})^2.
\end{equation*}
Witten suggested \cite{Witten} that $M_{Pl}=2\times10^{16}\;{\rm GeV}=M_U$, and used the large volume to explain the discrepancy between the Planck and unification scales.  Thus we expect
\begin{equation*}
{\rm ln}D_P\sim10^4.
\end{equation*}

To proceed, we recall how \cite{bfm} extracted particle states from the pixel algebra.  The point is simply that $N\times N+1$ matrices are precisely the spinor bundle over the fuzzy 2-sphere.  For finite $N$, we keep only a finite number, of order $N^2$ spinor spherical harmonics in the expansion of a section of this bundle.  Ignoring the compact dimensions, the pixel variables converge, as $N\rightarrow\infty$, to $\psi(\Omega )$, an operator valued measure on the spinor bundle.  These are the operators describing a single massless chiral super-particle in four dimensions, with fixed magnitude of the momentum and direction $\Omega$.  It is hoped that the incorporation of compact dimensions will allow us to generalize the particle content to include gravitons and gauge bosons.

In order to describe multi-particle states, as well as to obtain variable values of the longitudinal momentum, we introduce block diagonal $\psi$ matrices.  The size of an $M\times M+1$ block is interpreted as its momentum in units of $1/R$.  The usual permutation gauge symmetry of the space of block diagonal matrices, is interpreted as particle statistics, and the anti-commutation relations and spinor nature of the $\psi$ operators enforces the right spin statistics connection.

One must make a compromise between the number of particles allowed, and the number of spherical harmonics allowed in the momentum space wave function of a given particle (there must be many if it is to be localizable on the holographic screen\footnote{In experimental particle physics language this is localization in the detector.}).  The compromise which leads to the maximal particle entropy is to take blocks of size $M\sim N^{1/2}$.  This picture of the typical particle momentum and multiplicity which maximizes the entropy in dS space, can be derived in field theory by maximizing the entropy subject to the constraint that no black holes with radius of order the cosmological horizon are formed.

The super-Poincar\'{e} algebra arises in this formalism only as $N\rightarrow\infty$ and only for localizable particle states.  Corrections to the algebra should then scale like $N^{-1/2}$.  In particular, the commutator of the Poincar\'{e} Hamiltonian, $P_0$ and the supercharges $Q_a$ should be of order $N^{-1/2}M_PS_a$ where $S_a$ is an operator with matrix elements of order one.  It follows that the gravitino mass is given by a formula
\begin{equation*}
m_{3/2}=N^{-1/2}KM_P,
\end{equation*}
with $K$ of order one.  Our entropy formula gives
\begin{equation*}
\frac{3}{8}\frac{M_P^4}{\Lambda}=\pi(RM_P)^2=N^2{\rm ln}D_P\approx10^4N^2.
\end{equation*}
Comparing these two formulae, we get the one used in the text by lumping a factor $(\frac{8}{3})^{1/4}$ into $K$.

It should be noted that the Lorentz group arises in this formalism as the conformal group of the sphere.  The formalism is exactly rotation invariant for any $N$ and conformal transformations corresponding to boosts of moderate rapidity should not be affected much by restricting the space of functions on the sphere to the first $10^{30}$ spherical harmonics.

\section{Non-thermal dark matter}

This appendix recalls the non-thermal dark matter production scenario described in \cite{bmo}.  We assume that a particle $X$ with $m_X\gg m_{\cal B}$ decays, with a reheat temperature $T_{RH}<\Lambda_3$.  This produces an initial abundance of heavy pyrma-baryons
\begin{equation*}
Y_0=10^{-2}\frac{T_{RH}}{m_{{\cal B}}}.
\end{equation*}
$Y_0$ is, as usual the number density to entropy density ratio.  The first factor in this equation is simply the branching ratio that would appear for a massless pyrma-baryon, while the second suppression factor takes into account the fact that the mass is above the typical energy of decay products after thermalization.  The decay is relatively quick, so we can neglect annihilation of pyrma-baryons during the decay process.

Below $T_{RH}$ the pyrma-baryon abundance satisfies a Boltzmann equation driven only by annihilation.  Processes which create more pyrma-baryons have already fallen out of equilibrium.

We have
\begin{equation*}
\frac{dY}{dx}=-k\frac{Y^2}{x^{5/2}},
\end{equation*}
where $x=m_{\cal B}/T$, and
\begin{equation*}
k=\frac{2\pi m_{\cal B}m_{P}\sigma_0g_{* s}}{75g_*^{1/2}}\approx(1.4\times10^{15}\;{\rm TeV})m_{\cal B}\sigma_0.
\end{equation*}
$g_*$ is the number of massless degrees of freedom into which the pyrma-baryons annihilate and $g_{*s}$ the number that contribute to the entropy.  We have, in the last expression for $k$, approximated both of these by an average value of $50$ and written all remaining dimensionful quantities in TeV units.

The solution for the present day abundance is
\begin{equation*}
Y_f^{-1}=Y_i^{-1}+\frac{2k}{3}(x_i^{-3/2}-x_f^{-3/2}).
\end{equation*}
The last term is negligible, and so is the first if $T_{RH}$ is high enough for nucleosynthesis to occur in a normal fashion.  Thus
\begin{equation*}
Y_f=\frac{10^{-15}m_{\cal B}^{1/2}T_{RH}^{-3/2}\sigma_0^{-1}}{{\rm TeV}}.
\end{equation*}
The observed dark matter density is obtained if
\begin{equation*}
Y_f\frac{m_{\cal B}}{{\rm TeV}}=4.4\times10^{-13},
\end{equation*}
so we must have
\begin{equation*}
\Lambda_3>T_{RH}=0.017\frac{m_{\cal B}}{[\sigma_0\;({\rm TeV})^2]^{2/3}}\approx0.15\;m_{\cal B}/A^{2/3}.
\end{equation*}
In the last approximate equality we have taken $\Lambda_3\sim5$ TeV and used $A$ as the value of $\sigma_0\Lambda_3^2$.  This can be satisfied for heavy pyrma-baryon masses less than
\begin{equation*}
m_{\cal B}<6.7A^{2/3}\Lambda_3.
\end{equation*}
Recalling that $m_{{\cal B}_3}$ cannot be much bigger than $3\Lambda_3$ (in order to satisfy the bounds on the gluino mass) and that $A\sim1$, we are able to fit the observed dark matter abundance, the gross features of the dark matter signals in ATIC, PAMELA and PPB-BETS, as well as supersymmetric phenomenology.  The parameters of our model are tightly constrained by all of this data.

\section{Some computations}

In this appendix we present some computations related to the meta-stable state.  Below the scale $\Lambda_3$, the relevant superpotential is given by
\begin{equation*}
W=\sum_{i=1,3}(m_i+y_iS){\cal T}_i\bar{\cal T}_i+(m_2+y_2S)\Lambda_3\,{\rm tr}\,M+g_\mu SH_uH_d+\frac{g_T}{3}S^3+\ldots
\end{equation*}
where the pyrmeson ${\cal T}_2\bar{\cal T}_2=\Lambda_3M$ satisfies the usual quantum moduli space constraint $\det{M}-\Lambda_3P\tilde{P}=\Lambda_3^3$ and the heavy trianons ${\cal T}_{i=1,3}$ and $\bar{\cal T}_{i=1,3}$ have to be integrated out.  Parametrizing the pyrmeson and the pyrma-baryons as
\begin{equation*}
M=Z_a\lambda^a,\;\;\;\;\;\;P=i\Lambda_3e^{(q+p)/\Lambda_3},\;\;\;\;\;\;\tilde{P}=i\Lambda_3e^{(q-p)/\Lambda_3}
\end{equation*}
where $Z_0\equiv Z$, $\lambda^0=\sqrt{\frac{2}{3}}\,I$ and $\lambda^{a=1,\ldots,8}$ are Gell-Mann matrices, the quantum moduli space constraint can be satisfied for any $Z_a$ and $p$ by fixing $q$.  If $q\neq-\infty$ then $p$ is the NGB of the broken $U(1)_{{\cal B}_2}$.  The superpotential in terms of the unconstrained fields is simply
\begin{equation*}
W=\sum_{i=1,3}(m_i+y_iS){\cal T}_i\bar{{\cal T}}_i+\sqrt{6}\Lambda_3(m_2+y_2S)Z+g_\mu SH_uH_d+\frac{g_T}{3}S^3+\ldots
\end{equation*}
where again the heavy trianons have to be integrated out.  The F-terms are
\begin{eqnarray*}
-F_S^\dagger &=& \sum_{i=1,3}y_i{\cal T}_i\bar{{\cal T}}_i+\sqrt{6}y_2\Lambda_3Z+g_\mu H_uH_d+g_TS^2\\
-F_{H_u}^\dagger &=& g_\mu SH_d\\
-F_{H_d}^\dagger &=& g_\mu SH_u\\
-F_Z^\dagger &=& \sqrt{6}\Lambda_3(m_2+y_2S)\\
-F_{{\cal T}_i}^\dagger &=& (m_i+y_iS)\bar{{\cal T}}_i\\
-F_{\bar{{\cal T}}_i}^\dagger &=& (m_i+y_iS){\cal T}_i
\end{eqnarray*}
and there is a SUSY vacuum at $S_{{\rm SUSY}}=-\frac{m_2}{y_2}$ and $Z_{{\rm SUSY}}=-\frac{g_Tm_2^2}{\sqrt{6}y_2^3\Lambda_3}$ with all other VEVs to zero.  The potential, after integrating out the heavy trianons, has three contributions,
\begin{equation*}
V=V_{{\rm F-terms}}+V_{{\rm 1-loop}}+V_{{\rm D-terms}}
\end{equation*}
where
\begin{eqnarray*}
V_{{\rm F-terms}} &=& |\sqrt{6}y_2\Lambda_3Z+g_\mu H_uH_d+g_TS^2|^2+|g_\mu S|^2(|H_u|^2+|H_d|^2)\\
 && +((\partial^2K)^{-1}_{Z^\dagger Z})6|\Lambda_3|^2|m_2+y_2S|^2\nonumber\\
V_{{\rm 1-loop}} &=& \frac{9}{32\pi^2}\sum_{i=1,3}\sum_{\sigma=\pm1}\left[m_B^4\log\frac{m_B^2}{\Lambda^2}-m_F^4\log\frac{m_F^2}{\Lambda^2}\right]\\
V_{{\rm D-terms}} &=& \frac{1}{8}(g_1^2+g_2^2)(|H_u|^2-|H_d|^2)^2+\frac{1}{2}g_2^2|H_u^+H_d^{0\dagger}+H_u^0H_d^{-\dagger}|^2
\end{eqnarray*}
and the fermionic and bosonic masses for the heavy trianons are given by
\begin{eqnarray*}
m_F^2 &=& |m_i+y_iS|^2\\
m_B^2 &=& |m_i+y_iS|^2+\sigma|y_i(\sqrt{6}y_2\Lambda_3Z+g_\mu H_uH_d+g_TS^2)|.
\end{eqnarray*}
There are critical points of the potential with $H_u^+=H_d^-=0$.  Assuming this, the potential becomes invariant under $H_u^0\leftrightarrow H_d^0$ and thus there are critical points with $H_u^0=H_d^0$.  At critical points like these, the D-term contribution vanishes and the potential simplifies greatly.  The existence of a SUSY violating minimum is encoded in the strong dynamics of the $SU_P(3)$ gauge group and is therefore difficult to determine.

\section*{Acknowledgments}

We would like to thank Michael Dine, Nima Arkani-Hamed and Scott Thomas, for important conversations.  This research was supported in part by DOE grant number DE-FG03-92ER40689.

\end{document}